\def\techreport{}
\newcommand{\CLASSINPUTinnersidemargin}{1in}
\newcommand{\CLASSINPUTtoptextmargin}{1in}
\documentclass[12pt,draftclsnofoot,journal,onecolumn]{IEEEtran}
\usepackage{cite}
\ifCLASSINFOpdf
\else
\fi
\usepackage[dvips]{graphicx}
\DeclareGraphicsExtensions{.pdf,.jpeg,.png}

\usepackage[cmex10]{amsmath}
\usepackage{amssymb}
\usepackage{amsthm}

\usepackage{algorithmic}
\usepackage{array}

\usepackage{mdwmath}
\usepackage{mdwtab}
\usepackage[caption=false,font=footnotesize]{subfig}
\usepackage{fixltx2e}
\usepackage{url}
\ifdefined\flagendfloatboulat
\usepackage[nolists,nomarkers]{endfloat}

\let\MYoriglatexcaption\caption
\renewcommand{\caption}[2][\relax]{\MYoriglatexcaption[#2]{#2}}
\fi
\newtheorem{theorem}{Theorem}
\newtheorem*{theorem*}{Theorem}

\newtheorem{fact}{Fact}
\newtheorem{lemma}{Lemma}

\newtheorem{definition}{Definition}

\newcounter{customtheoremcounter}
\newtheorem{customtheorem}{Theorem}[customtheoremcounter]

\newcommand{\expected}[1]{\mathbb{E}\left[#1\right]}
\newcommand{\prob}[1]{\mathbb{P}\left(#1\right)}
\DeclareMathOperator{\sinc}{sinc}
\DeclareMathOperator{\sech}{sech}
\DeclareMathOperator{\Var}{Var}

\newcommand{\plusser}[2] {
        \put(#1,#2){\begin{picture}(5,5)
                    \put(0,0) {\circle{5}}
                    \put(0,-2) {\line(0,1) {4}}
                    \put(-2,0){\line(1,0) {4}}
                    \end{picture}}
                        }

\begin{document}
\title{Limits of Reliable Communication with Low Probability of Detection on AWGN Channels}

\author{Boulat~A.~Bash,~%
        Dennis~Goeckel,~%
        Don~Towsley%
\thanks{B. A. Bash and D. Towsley are with the School of Computer Science, University of Massachusetts, Amherst, Massachusetts.}%
\thanks{D. Goeckel is with the Electrical and Computer Engineering Department, University of Massachusetts, Amherst, Massachusetts.}%
\thanks{This research was sponsored by the National Science Foundation under 
  grants CNS-0905349 and CNS-1018464, and by the U.S. Army Research Laboratory 
  and the U.K. Ministry of Defence under Agreement Number W911NF-06-3-0001. 
  The views and conclusions contained in this document are those of the 
  author(s) and should not be interpreted as representing the official 
  policies, either expressed or implied, of the U.S. Army Research Laboratory, 
  the U.S. Government, the U.K. Ministry of Defence or the U.K.  Government. 
  The U.S. and U.K. Governments are authorized to reproduce and distribute 
  reprints for Government purposes notwithstanding any copyright notation 
  hereon.}}
\maketitle
\vspace{-0.30in}
\begin{abstract}
We present a square root limit on the amount of information transmitted reliably and with low probability of detection (LPD) over additive white Gaussian noise (AWGN) channels. Specifically, if the transmitter has AWGN channels to an intended receiver and a warden, both with non-zero noise power, we prove that $o(\sqrt{n})$ bits can be sent from the transmitter to the receiver in $n$ channel uses while lower-bounding 
$\alpha+\beta\geq1-\epsilon$ for any $\epsilon>0$, where $\alpha$ and $\beta$ respectively denote the warden's probabilities of a false alarm when the sender is not transmitting and a missed detection when the sender is transmitting.
Moreover, in most practical scenarios, a lower bound on the noise power on the channel between the transmitter and the warden is known and $\mathcal{O}(\sqrt{n})$ bits can be sent in $n$ LPD channel uses. Conversely, attempting to transmit more than $\mathcal{O}(\sqrt{n})$ bits either results in detection by the warden with probability one or a non-zero probability of decoding error at the receiver as $n\rightarrow\infty$.

\end{abstract}
\IEEEpeerreviewmaketitle

\section{Introduction}
Securing information transmitted over wireless links is of paramount concern
  for consumer, industrial, and military applications.
Typically data transmitted in wireless networks is secured
  from interception by an eavesdropper using various
  encryption and key exchange protocols.
However, there are many real-life scenarios where standard
  cryptographic security is not sufficient.
Encrypted data arouses suspicion, and even the most theoretically robust 
  encryption can often be defeated by a determined adversary using 
  non-computational methods such as side-channel analysis.
Such scenarios require \emph{low probability of detection} (LPD) communication
  which prevents the detection of transmissions in the first place.

While practical LPD communications has been studied by the spread-spectrum
  community \cite[Pt.~5, Ch.~1]{simon94ssh}, 
  \cite[Ch.~1.4 and 14]{varakin85ss}, the information-theoretic
  limits have not been explored.
We thus develop fundamental bounds on LPD communication over 
  wireless channels subject to additive white Gaussian noise (AWGN).
In our scenario, Alice communicates with Bob over an AWGN channel, while 
  passive eavesdropper Warden Willie attempts to detect her transmission.
The channel between Alice and Willie is also AWGN and
  Willie is passive in that he does not actively jam Alice's 
  channel.
Alice transmits low-power signals to Bob that Willie attempts to 
  classify as either noise on his channel from Alice or Alice's signals to Bob.
If he detects communication, Willie can potentially shut
  the channel down or otherwise punish Alice.
If the noise on the channel between Willie and Alice has non-zero 
  power, Alice can communicate
  with Bob while tolerating a certain probability of detection, which she can 
  drive down by transmitting with low enough power.
Thus, Alice potentially transmits non-zero mutual information across the LPD
  channel to Bob in $n$ uses of the channel.

Our problem is related to imperfect steganography, which considers hiding 
  information by altering the 
  properties of fixed-size, finite-alphabet covertext objects 
  (such as images or software binary code) while
  tolerating some fixed probability of detection 
  of hidden information by the warden.
The square root law of steganography in the passive warden environment states 
  that $\mathcal{O}(\sqrt{n})$
  symbols in covertext of size $n$ may safely be modified to hide
  an $\mathcal{O}(\sqrt{n}\log{n})$-bit steganographic message
  \cite[Ch.~13]{fridrich09stego},
  where the $\log{n}$ factor stems directly from the fact that transmission
  to Bob is noiseless \cite[Ch.~8]{fridrich09stego}.
In our scenario, Alice uses the noise on her channel to Willie instead of the 
  statistical properties of
  the covertext to hide information.
However, having to code against the noise on her channel to Bob allows only 
  $\mathcal{O}(\sqrt{n})$
  bits to be sent in $n$ uses of the LPD channel.\footnote{The 
  amount of information that could be transmitted by Alice to Bob using
  a \emph{noiseless} LPD channel would be infinite due to it being 
  continuously-valued, and a noiseless channel between Alice and Willie
  would preclude the existence of an LPD channel between Alice and Bob.}
The mathematics of statistical 
  hypothesis testing yields a square root law in both problems, but as
  answers to different questions due 
  to the fundamental differences in the communication channels.
This relationship is discussed further at the end of Section 
  \ref{sec:achievability}.

We state our main result that limits the amount of information that can
  be transmitted on the LPD channel 
  between Alice and Bob %
  using asymptotic notation \cite[Ch.~3.1]{clrs2e}
  where $f(n)=\mathcal{O}(g(n))$ denotes an asymptotically tight 
  upper bound on $f(n)$ (i.e.~there exist constants $m,n_0>0$ such 
  that $0\leq f(n)\leq m g(n)$ for all $n\geq n_0$), 
  $f(n)=o(g(n))$ denotes an upper bound on $f(n)$ that is not 
  asymptotically tight (i.e.~for any constant $m>0$, there exists constant 
  $n_0>0$ such that $0\leq f(n)<m g(n)$ for all $n\geq n_0$),
  and $f(n)=\omega(g(n))$ denotes a lower bound on $f(n)$ that is not 
  asymptotically tight (i.e.~for any constant $m>0$, there exists constant 
  $n_0>0$ such that $0\leq m g(n)<f(n)$ for all $n\geq n_0$):
\begin{theorem*}[Square root law]\label{th:main}
Suppose the channels between Alice and each of Bob and Willie experience 
  additive white Gaussian noise (AWGN) with powers 
  $\sigma_b^2>0$ and $\sigma_w^2>0$, respectively, where $\sigma_b^2$ and 
  $\sigma_w^2$ are constants.
Denote by $\alpha$ the probability that Willie raises a false alarm when Alice
  is not transmitting, and by $\beta$ the probability that Willie
  does not detect a transmission by Alice.
Then, provided that Alice and Bob have a shared secret of sufficient length,
  for any $\epsilon > 0$ and unknown $\sigma_w^2$, 
  Alice can reliably 
  (i.e.~with arbitrary low probability of decoding error)
  transmit $o(\sqrt{n})$ information
  bits to Bob in $n$ channel uses while lower-bounding Willie's sum
  of the probabilities of detection errors $\alpha+\beta\geq1-\epsilon$.
Moreover, if Alice knows a lower bound $\hat{\sigma}_w^2>0$ to the power of 
  the AWGN 
  on Willie's channel $\sigma_w^2$ (i.e.~$\sigma_w^2\geq\hat{\sigma}_w^2$), 
  she can transmit $\mathcal{O}(\sqrt{n})$ bits in $n$ channel uses while
  maintaining the lower bound $\alpha+\beta\geq1-\epsilon$.
Conversely, if Alice attempts to transmit $\omega(\sqrt{n})$ bits in $n$ 
  channel uses, then, as $n\rightarrow \infty$, either Willie detects her 
  with arbitrarily low probability of error or Bob cannot decode her message
  reliably, regardless of the length of the shared secret. 
\end{theorem*}

To enable LPD communication, Alice and Bob possess
  a common secret randomness resource.
While in the information-theoretic analysis of encrypted communication such 
  a resource is a one-time pad \cite{shannon49sec}, in the construction of our
  proofs it is a secret codebook that is shared between Alice and Bob
  prior to communication and which is the only component of their
  system that is unknown to Willie.
This follows ``best practices'' in security system design
  as the security of the LPD communication system depends only on the shared  
  secret \cite{menezes96HAC}.
We also note that, since LPD communication allows transmission of 
  $\mathcal{O}(\sqrt{n})$ bits in $n$ channel uses and, considering
  $\lim_{n\rightarrow\infty}\frac{\mathcal{O}(\sqrt{n})}{n}=0$,
  the information-theoretic capacity of the LPD channel is zero, unlike
  many other communications settings where it is a positive constant.
However, a significant amount of information can still be transmitted using
  this channel. %
We are thus concerned with the number of information bits
  transmitted in $n$ channel uses, as opposed to the bits per channel use.

After introducing our channel model and hypothesis testing 
  background in Section 
  \ref{sec:prerequisites}, we prove the achievability of the square root law
  in Section \ref{sec:achievability}.
We then prove the converse in Section \ref{sec:converse}.
We discuss the relationship to previous work, the impact of Willie's prior
  knowledge of Alice's transmission state, and the mapping to the
  continuous-time channel in Section \ref{sec:discussion}, and conclude in
  Section \ref{sec:conclusion}.

\section{Prerequisites}
\label{sec:prerequisites}

\subsection{Channel Model}
\begin{figure}[h]
\begin{center}
\begin{picture}(85,42)
\put(33,38){\framebox(14,4){\parbox{14\unitlength}{\small \centering secret}}}
\put(7,40){\vector(0,-1){7}}
\put(73,40){\vector(0,-1){7}}
\put(7,40){\line(1,0){26}}
\put(47,40){\line(1,0){26}}
\put(0,27){\framebox(15,6){Alice}}
\put(15,30){\vector(1,0){31.25}}
\put(15,30){\makebox(25,5){\small $f_1,f_2,\ldots,f_n$}}
\put(40,30){\circle*{1}}
\put(40,30){\vector(0,-1){12.5}}
\plusser{40}{15}
\put(42.5,15){\vector(1,0){12.5}}
\put(55,12){\framebox(15,6){Willie}}
\put(32,0){\makebox(53,12){\parbox{53\unitlength}{\small \centering decide $z_1^{(w)},z_2^{(w)},\ldots,z_n^{(w)}$ or\\ $f_1+z_1^{(w)},f_2+z_2^{(w)},\ldots,f_n+z_n^{(w)}$?}}}
\put(25,15){\vector(1,0){12.5}}
\put(18,12){\makebox(7,6){$z_i^{(w)}$}}
\put(45.25,17.5){\makebox(7,6){$z_i^{(b)}$}}
\plusser{48.75}{30}
\put(48.75,23.5){\vector(0,1){4}}
\put(51.25,30){\vector(1,0){13.75}}
\put(65,27){\framebox(15,6){Bob}}
\put(55,22){\makebox(30,5){\small decode $f_1,f_2,\ldots,f_n$}}
\end{picture}
\end{center}
\vspace{-0.20in}
\caption{System framework: Alice and Bob share a secret before the transmission.
  Alice encodes information into a vector of real 
  symbols $\mathbf{f}=\{f_i\}_{i=1}^n$ and transmits it on an AWGN 
  channel to Bob, while Willie attempts to classify his vector of observations 
  of the  channel from Alice $\mathbf{y}_w$ as either an AWGN vector
  $\mathbf{z}_w=\{z^{(w)}_i\}_{i=1}^n$ or a vector
  $\{f_i+z^{(w)}_i\}_{i=1}^n$ of transmissions corrupted by AWGN.}
\label{fig:framework}
\vspace{-0.10in}
\end{figure}
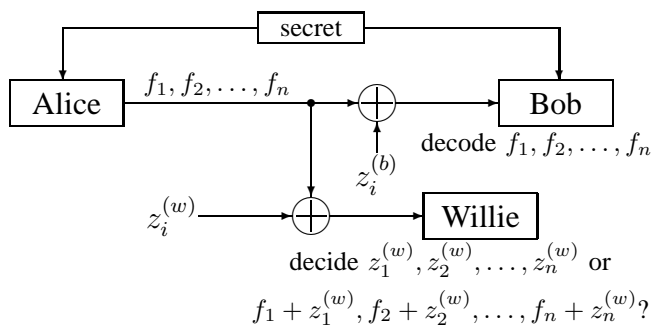

We use the discrete-time AWGN channel model with real-valued symbols (and defer
  discussion of the mapping to a continuous-time channel
  to Section \ref{sec:ct_channel}).
Our formal system framework is depicted in Figure \ref{fig:framework}.
Alice transmits a vector of $n$ real-valued symbols 
  $\mathbf{f}=\{f_i\}_{i=1}^n$.
Bob receives vector $\mathbf{y}_b=\{y^{(b)}_i\}_{i=1}^n$ where
  $y^{(b)}_i=f_i+z^{(b)}_i$ with an independent and identically distributed 
  (i.i.d.) $z^{(b)}_i\sim \mathcal{N}(0,\sigma_b^2)$.
Willie observes vector $\mathbf{y}_w=\{y^{(w)}_i\}_{i=1}^n$ where 
  $y^{(w)}_i=f_i+z^{(w)}_i$, with 
  i.i.d.~$z^{(w)}_i\sim\mathcal{N}(0,\sigma_w^2)$.
Willie uses statistical hypothesis tests on $\mathbf{y}_w$ to determine 
  whether Alice is communicating, which we discuss next.

\subsection{Hypothesis Testing}
Willie expects vector $\mathbf{y}_w$ of $n$ channel readings to be 
  consistent with his channel noise model.
He performs a statistical hypothesis test on this vector, with the null 
  hypothesis $H_0$ being that Alice is not communicating.
In this case each sample is i.i.d.~$y^{(w)}_i\sim \mathcal{N}(0,\sigma_w^2)$.
The alternate hypothesis $H_1$ is that Alice is transmitting,
  which corresponds to samples $y^{(w)}_i$ coming from a different distribution.
Willie can tolerate some false positives, or cases when his statistical test 
  incorrectly accuses Alice.
This rejection of $H_0$ when it is true is known as the type I error 
  (or false alarm), and, following the standard nomenclature,
  we denote its probability by $\alpha$ \cite{lehmann05stathyp}.
Willie's test may also miss Alice's transmissions.
Acceptance of $H_0$ when it is false is known as the type II error (or missed 
  detection), and we denote its probability by $\beta$.
We assume that Willie uses classical hypothesis testing with equal prior
  probabilities of each hypothesis being true (and discuss the generalization 
  to unequal prior probabilities in Section \ref{sec:bayes}).
Then, the lower bound on the sum $\alpha+\beta$ characterizes the necessary 
  trade-off between the false alarms and the missed detections in the design of
  a hypothesis test.

\section{Achievability of Square Root Law}
\label{sec:achievability}

Willie's objective is to determine whether Alice
  transmits given the vector of observations 
  $\mathbf{y}_w$ of his channel from Alice.
Denote the probability distribution of Willie's channel observations 
  when Alice does not transmit (i.e.~when  
  $H_0$ is true) as $\mathbb{P}_0$, and the probability distribution of the 
  observations when Alice transmits (i.e.~when $H_1$ is true) as 
  $\mathbb{P}_1$.
To strengthen the achievability result, 
  we assume that Alice's channel input distribution, as well as the distribution
  of the AWGN on the channel between Alice and Willie, are known to Willie.
Then $\mathbb{P}_0$ and $\mathbb{P}_1$ are known to Willie, and he can 
  construct an optimal statistical hypothesis test (such as the Neyman--Pearson 
  test) that minimizes the sum of error probabilities 
  $\alpha+\beta$ \cite[Ch.~13]{lehmann05stathyp}.
The following holds for such a test:

\begin{fact}[Theorem 13.1.1 in \cite{lehmann05stathyp}]
\label{fact:totvar_testperf}
For the optimal test, 
\begin{eqnarray*}
\alpha+\beta&=&1-\mathcal{V}_T(\mathbb{P}_0,\mathbb{P}_1)
\end{eqnarray*}
\end{fact}
\noindent where $\mathcal{V}_T(\mathbb{P}_0,\mathbb{P}_1)$ is the total variation distance 
  between $\mathbb{P}_0$ and $\mathbb{P}_1$
  defined as follows:
\begin{definition}[Total variation distance \cite{lehmann05stathyp}]
The \emph{total variation distance} between two continuous probability measures 
  $\mathbb{P}_0$ and $\mathbb{P}_1$ is
\begin{eqnarray}
\label{eq:totvardef}\mathcal{V}_T(\mathbb{P}_0,\mathbb{P}_1)&=&\frac{1}{2}\|p_0(x)-p_1(x)\|_1
\end{eqnarray}
\noindent where $p_0(x)$ and $p_1(x)$ are densities
  of $\mathbb{P}_0$ and $\mathbb{P}_1$, respectively, and 
  $\|a-b\|_1$ is the $\mathcal{L}_1$ norm.
\end{definition}
Implicit in the above is that the \emph{a priori} probabilities of $H_0$ and
  $H_1$ are unknown to Willie.
We discuss the inclusion of knowledge of prior probabilities in 
  Section \ref{sec:bayes}. %

Since total variation lower-bounds the error of all 
  hypothesis tests Willie can use, a clever choice of $\mathbf{f}$ allows Alice
  to limit Willie's detector performance.
Unfortunately, the total variation metric is unwieldy for products 
  of probability measures, which are used in the analysis of the vectors of 
  observations.
We thus use Pinsker's inequality:

\begin{fact}[Pinsker's inequality (Lemma 11.6.1 in \cite{cover02IT})]
\label{fact:pinsker}
\begin{eqnarray*}
\mathcal{V}_T(\mathbb{P}_0,\mathbb{P}_1)\leq\sqrt{\frac{1}{2}\mathcal{D}(\mathbb{P}_0\|\mathbb{P}_1)}
\end{eqnarray*}
\end{fact}
\noindent where relative entropy $\mathcal{D}(\mathbb{P}_0\|\mathbb{P}_1)$ is 
  defined as follows:

\begin{definition}
The \emph{relative entropy} (also known as \emph{Kullback--Leibler divergence})
  between two probability measures 
  $\mathbb{P}_0$ and $\mathbb{P}_1$ is:
\begin{eqnarray}
\label{eq:rel_entropy}
\mathcal{D}(\mathbb{P}_0\|\mathbb{P}_1)&=&\int_{\mathcal{X}}p_0(x)\ln\frac{p_0(x)}{p_1(x)}dx
\end{eqnarray}
\noindent where $\mathcal{X}$ is the support of $p_1(x)$.
\end{definition}

If $\mathbb{P}^n$ is the distribution of a sequence $\left\{X_i\right\}_{i=1}^n$
  where each $X_i\sim\mathbb{P}$ is i.i.d., then:

\begin{fact}[Relative entropy product]\label{fact:rel_ent_product} From the chain rule for relative entropy \cite[Eq.~(2.67)]{cover02IT}:
\begin{eqnarray*}
\mathcal{D}(\mathbb{P}_0^n\|\mathbb{P}_1^n) &=&n \mathcal{D}(\mathbb{P}_0\|\mathbb{P}_1)
\end{eqnarray*}
\end{fact}

Relative entropy is directly related to Neyman--Pearson hypothesis testing
  via the Chernoff--Stein Lemma \cite[Ch.~11.8]{cover02IT}:
  for a given $\alpha<\nu$ with $0<\nu<\frac{1}{2}$,
  $\lim_{\nu\rightarrow0}\lim_{n\rightarrow\infty}\frac{1}{n}\ln\beta^*=-\mathcal{D}(\mathbb{P}_0\|\mathbb{P}_1)$ where $\beta^*=\min\beta$.
Thus, upper-bounding the relative entropy limits the performance of
  the Neyman--Pearson hypothesis test.
Indeed, the steganography community often concludes their proofs 
  by showing an upper bound on the relative entropy
  \cite{cachin04itstego, fridrich09stego}.
However, we take the extra step of lower-bounding $\alpha+\beta$ since it has a 
  natural signal processing 
  interpretation via the receiver operating characteristic (ROC) curve 
  \cite[Ch.~2.2.2]{vantrees01part1}, 
  which plots probability of detection $1-\beta$ versus $\alpha$.
Since $1-\beta\geq \alpha$ and $\alpha+\beta\geq 1-\epsilon$, small $\epsilon$
  implies that the ROC curve lies very close to the line of no-discrimination
  (the diagonal line where $1-\beta=\alpha$) over the entire domain of $\alpha$
  because $\alpha+\epsilon\geq 1-\beta\geq\alpha$.

We use Taylor's theorem with the Lagrange form of the remainder to 
  upper-bound the relative entropy, and here we restate it as a lemma.
\begin{lemma}[Taylor's theorem with the remainder]\label{lemma:taylor}
If $f(x)$ is a function with $n+1$ continuous derivatives on the interval 
  $[u,v]$, then
\begin{eqnarray}
f(v)&=&f(u)+f'(u)(v-u)+\ldots+\frac{f^{(n)}(u)}{n!}(v-u)^{n}+\frac{f^{(n+1)}(\xi)}{(n+1)!}(v-u)^{n+1}
\end{eqnarray}
where $f^{(n)}(x)$ denotes the $n^{\text{th}}$ derivative of $f(x)$, and $\xi$
  satisfies $u\leq \xi\leq v$.
\end{lemma}
The proof can be found in, e.g.~\cite[Ch.~V.3]{lang97ugradanalysis}.
Note that if the remainder term is negative on $[u,v]$, then the sum of the
  zeroth through $n^{\text{th}}$ order terms yields an upper bound on $f(v)$.

We now state the achievability theorem under an average power constraint:

\begin{customtheorem}[Achievability]
\label{th:achievability}
Suppose Willie's channel is subject to AWGN with average power $\sigma^2_w>0$
  and suppose that Alice and Bob share a secret of sufficient length.
Then Alice can maintain Willie's sum of the probabilities of 
  detection errors $\alpha+\beta\geq 1-\epsilon$ for any $\epsilon>0$
  while reliably transmitting $o(\sqrt{n})$ bits to Bob
  over $n$ uses of an AWGN channel when $\sigma_w^2$ is 
  unknown and $\mathcal{O}(\sqrt{n})$ bits over $n$ channel uses if she knows
  a lower bound $\sigma_w^2\geq\hat{\sigma}_w^2$ for some 
  $\hat{\sigma}_w^2>0$.
\end{customtheorem}

\begin{IEEEproof}
\textbf{Construction:}
Alice's channel encoder takes as input blocks of length $M$ bits and encodes
  them into codewords of length $n$ at the rate of $R=M/n$ bits/symbol.
We employ random coding arguments and independently generate $2^{nR}$
  codewords $\{\mathbf{c}(W_k), k=1,2,\ldots,2^{nR}\}$ from $\mathbb{R}^n$ 
  for messages $\{W_k\}_{k=1}^{2^{nR}}$, each according 
  to $p_{\mathbf{X}}(\mathbf{x})=\prod_{i=1}^{n}p_X(x_i)$, where
  $X\sim \mathcal{N}(0,P_f)$ and $P_f$ is defined later. %
The codebook is used only to send a single message and is the secret 
  not revealed to Willie, though
  he knows how it is constructed, including the value of $P_f$.
The size of this secret is discussed in the remark following
  the proof of Theorem \ref{th:achievability_peak_power}.

The channel between Alice and Willie is corrupted by AWGN with power 
  $\sigma_w^2$.
Willie applies statistical hypothesis testing on a vector of $n$ channel 
  readings $\mathbf{y}_w$ to decide whether Alice transmits.
Next we show how Alice can limit the performance of Willie's methods.

\textbf{Analysis:}
Consider the case when Alice transmits codeword $\mathbf{c}(W_k)$.
Suppose that Willie employs a detector that implements an optimal hypothesis 
  test on his $n$ channel readings.
His null hypothesis $H_0$ is that Alice does not transmit and that
  he observes noise on his channel.
His alternate hypothesis $H_1$ is that Alice transmits and that
  he observes Alice's codeword corrupted by noise.
By Fact \ref{fact:totvar_testperf}, the sum of the probabilities of Willie's
  detector's errors is expressed 
  by $\alpha+\beta=1-\mathcal{V}_T(\mathbb{P}_0,\mathbb{P}_1)$, where
  the total variation distance is
  between the distribution $\mathbb{P}_0$ of $n$ noise readings that Willie
  expects to observe under his null hypothesis and the distribution 
  $\mathbb{P}_1$ of the codeword transmitted by Alice corrupted 
  by noise.
Alice can lower-bound the sum of the error probabilities by
  upper-bounding the total variation distance:
  $\mathcal{V}_T(\mathbb{P}_0,\mathbb{P}_1)\leq\epsilon$.

The realizations of noise $z^{(w)}_i$ in vector $\mathbf{z}_w$  
  are zero-mean i.i.d.~Gaussian random
  variables with 
  variance $\sigma_w^2$, and, thus, $\mathbb{P}_0=\mathbb{P}_w^n$ where
  $\mathbb{P}_w=\mathcal{N}(0,\sigma_w^2)$.
Recall that Willie does not know the codebook.
Therefore, Willie's probability distribution of the transmitted symbols is
  of zero-mean i.i.d.~Gaussian random variables 
  with variance $P_f$.
Since noise is independent of the transmitted symbols, Willie
  observes vector $\mathbf{y}_w$, where 
  $y^{(w)}_i\sim\mathcal{N}(0,P_f+\sigma_w^2)=\mathbb{P}_s$ is 
  i.i.d., and thus, $\mathbb{P}_1=\mathbb{P}_s^n$.
By Facts \ref{fact:pinsker} and \ref{fact:rel_ent_product}:
\begin{eqnarray*}
\mathcal{V}_T(\mathbb{P}_w^n,\mathbb{P}_s^n)\leq\sqrt{\frac{1}{2}\mathcal{D}(\mathbb{P}_w^n\|\mathbb{P}_s^n)}=\sqrt{\frac{n}{2}\mathcal{D}(\mathbb{P}_w\|\mathbb{P}_s)}
\end{eqnarray*}
In our case the relative entropy is:
\begin{eqnarray*}
\label{eq:kl}
\mathcal{D}(\mathbb{P}_w\|\mathbb{P}_s)&=&\frac{1}{2}\left[\ln\left(1+\frac{P_f}{\sigma_w^2}\right)-\left(1+\left(\frac{P_f}{\sigma_w^2}\right)^{-1}\right)^{-1}\right]
\end{eqnarray*}

Since the first three derivatives of $\mathcal{D}(\mathbb{P}_w\|\mathbb{P}_s)$ 
  with respect to $P_f$ are continuous, we can apply Lemma \ref{lemma:taylor}.
The zeroth and first order terms of the Taylor series expansion with respect
  to $P_f$ around $P_f=0$ are zero.
However, the second order term is:
\begin{eqnarray*}
\frac{P_f^2}{2!}\times\left.\frac{\partial^2 \mathcal{D}(\mathbb{P}_w\|\mathbb{P}_s)}{\partial P_f^2}\right|_{P_f=0}&=&\frac{P_f^2}{4\sigma_w^4}
\end{eqnarray*}
That relative entropy is locally quadratic is well-known 
  \cite[Ch.~2.6]{kullback59IT}; in fact 
  $\left.\frac{\partial^2 \mathcal{D}(\mathbb{P}_w\|\mathbb{P}_s)}{\partial P_f^2}\right|_{P_f=0}=\frac{1}{2\sigma_w^4}$
  is the Fisher information that an observation of noise carries about its 
  power.
Now, the remainder term is:
\begin{eqnarray*}
\frac{P_f^3}{3!}\times\left.\frac{\partial^3 \mathcal{D}(\mathbb{P}_w\|\mathbb{P}_s)}{\partial P_f^3}\right|_{P_f=\xi}&=&\frac{P_f^3}{3!}\times\frac{\xi-2\sigma_w^2}{(\xi+\sigma_w^2)^4}
\end{eqnarray*}
where $\xi$ satisfies $0\leq\xi\leq P_f$.
Suppose Alice sets her average symbol power 
  $P_f\leq\frac{cf(n)}{\sqrt{n}}$, where $c=2\epsilon\sqrt{2}$ and 
  $f(n)=\mathcal{O}(1)$ is a function defined later.
Since the remainder is negative when $P_f<2\sigma_w^2$,
  for $n$ large enough, we can 
  upper-bound relative entropy with the second order term as follows:
\begin{eqnarray}
\label{eq:tv_taylor_bound}
\mathcal{V}_T(\mathbb{P}_w^n,\mathbb{P}_s^n)\leq \frac{P_f}{2\sigma_w^2}\sqrt{\frac{n}{2}}\leq\frac{\epsilon f(n)}{\sigma_w^2}
\end{eqnarray}

In most practical scenarios
  Alice knows a lower bound $\sigma_w^2\geq\hat{\sigma}_w^2$ and can set 
  $f(n)=\hat{\sigma}_w^2$ (a conservative
  lower bound is the thermal noise power of the best currently available
  receiver).
If $\sigma_w^2$ is unknown, Alice can set $f(n)$ such that $f(n)=o(1)$ and
  $f(n)=\omega(1/\sqrt{n})$ (the latter condition is needed to bound      
    Bob's decoding error probability).
In either case, Alice upper-bounds
  $\mathcal{V}_T(\mathbb{P}_w^n,\mathbb{P}_s^n)\leq \epsilon$,
  limiting the performance of Willie's detector.

Next we examine the probability $\mathbb{P}_e$ of Bob's decoding 
  error averaged over all possible codebooks.
Since Alice's symbol power $P_f$ is a decreasing function of the codeword 
  length $n$, the standard channel coding results for constant power 
  (and constant rate) do not directly apply.
Let Bob employ a maximum-likelihood (ML) decoder (i.e.~minimum distance decoder)
  to process the received vector $\mathbf{y}_b$ when $\mathbf{c}(W_k)$
  was sent.
The decoder suffers an error event $E_i(\mathbf{c}(W_k))$ when $\mathbf{y}_b$ 
  is closer to another codeword $\mathbf{c}(W_i)$, $i\neq k$.
The decoding error probability, averaged over all codebooks, is then:
\begin{eqnarray}
\mathbb{P}_e&=&\mathbb{E}_{\mathbf{c}(W_k)}\left[\mathbb{P}\left(\cup_{i=0,i\neq k}^{2^{nR}}E_i(\mathbf{c}(W_k))\right)\right]\nonumber\\
\label{eq:unionbound}&\leq&\mathbb{E}_{\mathbf{c}(W_k)}\left[\sum_{i=0,i\neq k}^{2^{nR}}\mathbb{P}\left(E_i(\mathbf{c}(W_k))\right)\right]\\
\label{eq:expectationthroughsum}&=&\sum_{i=0,i\neq k}^{2^{nR}}\mathbb{E}_{\mathbf{c}(W_k)}\left[\mathbb{P}\left(E_i(\mathbf{c}(W_k))\right)\right]
\end{eqnarray}
where $\mathbb{E}_X[\cdot]$ denotes the expectation operator over random 
  variable $X$ and (\ref{eq:unionbound}) follows from the union bound.
Let $\mathbf{d}=\mathbf{c}(W_k)-\mathbf{c}(W_i)$.
Then $\|\mathbf{d}\|_2$ is the distance between two codewords,
  where $\|\cdot\|_2$ is the $\mathcal{L}_2$ norm.
Since codewords are independent and Gaussian, 
  $d_j\sim\mathcal{N}(0,2P_f)$ for $j=1,2,\ldots,n$ and 
  $\|\mathbf{d}\|_2^2=2P_f U$, where $U\sim\chi^2_n$, with $\chi^2_n$ denoting 
  the chi-squared distribution with $n$ degrees of freedom.
Therefore, by \cite[Eq.~(3.44)]{madhow08digicom}:
\begin{eqnarray*}
\mathbb{E}_{\mathbf{c}(W_k)}\left[\mathbb{P}\left(E_i(\mathbf{c}(W_k))\right)\right]&=&\mathbb{E}_{U}\left[Q\left(\sqrt{\frac{P_f U}{2\sigma_b^2}}\right)\right]
\end{eqnarray*}
where $Q(x)=\frac{1}{\sqrt{2\pi}}\int_{x}^{\infty}e^{-t^2/2}dt$.
Since $Q(x)\leq\frac{1}{2}e^{-x^2/2}$ \cite[Eq.~(5)]{chiani03qfunction} and
  $P_f=\frac{cf(n)}{\sqrt{n}}$:
\begin{eqnarray}
\mathbb{E}_{U}\left[Q\left(\sqrt{\frac{P_f U}{2\sigma_b^2}}\right)\right]&\leq&\mathbb{E}_{U}\left[\exp\left(-\frac{cf(n) U}{4\sqrt{n}\sigma_b^2}\right)\right]\nonumber\\
\label{eq:gammatrick}&=&\int_0^\infty \frac{e^{-\frac{c f(n) u}{4\sqrt{n}\sigma_b^2}-\frac{u}{2}}2^{-\frac{n}{2}}u^{\frac{n}{2}-1}}{\Gamma(n/2)}du\\
\label{eq:singlecodeworderror}&=&2^{-n/2}\left(\frac{1}{2}+\frac{cf(n)}{4\sqrt{n}\sigma_b^2}\right)^{-n/2}
\end{eqnarray}
where (\ref{eq:singlecodeworderror}) is from the substitution
  $v=u\left(\frac{1}{2}+\frac{cf(n)}{4\sqrt{n}\sigma_b^2}\right)$ in 
  \eqref{eq:gammatrick} and the
  definition of the Gamma function $\Gamma(n)=\int_0^\infty x^{n-1}e^{-x}dx$.
Since $\frac{1}{2}+\frac{cf(n)}{4\sqrt{n}\sigma_b^2}=2^{\log_2\left(\frac{1}{2}+\frac{cf(n)}{4\sqrt{n}\sigma_b^2}\right)}$:
\begin{eqnarray*}
\mathbb{E}_{\mathbf{c}(W_k)}\left[\mathbb{P}\left(E_i(\mathbf{c}(W_k))\right)\right]&\leq&2^{-\frac{n}{2}\log_2\left(1+\frac{cf(n)}{2\sqrt{n}\sigma_b^2}\right)}
\end{eqnarray*}
  for all $i$, and (\ref{eq:expectationthroughsum}) becomes:
\begin{eqnarray}
\mathbb{P}_e&\leq&2^{nR-\frac{n}{2}\log_2\left(1+\frac{cf(n)}{2\sqrt{n}\sigma_b^2}\right)}
\end{eqnarray}
Since $f(n)=\omega(1/\sqrt{n})$, if rate 
  $R=\frac{\rho}{2}\log_2\left(1+\frac{cf(n)}{2\sqrt{n}\sigma_b^2}\right)$
  for a constant $\rho<1$, as $n$ increases,
  the probability of Bob's decoding error averaged over all codebooks decays 
  exponentially to zero 
  and Bob obtains $nR=n\frac{\rho}{2}\log_2\left(1+\frac{cf(n)}{2\sqrt{n}\sigma_b^2}\right)$ LPD bits in $n$ channel uses. 
Since $\ln(1+x)\leq x$ with equality when $x=0$, 
  $nR\leq\frac{\sqrt{n}\rho cf(n)}{4\sigma_b^2\ln 2}$, approaching equality
  as $n$ gets large.
Thus, Bob receives $o(\sqrt{n})$ bits in $n$ channel 
  uses, and $\mathcal{O}(\sqrt{n})$ bits in $n$ channel uses if 
  $f(n)=\hat{\sigma}_w^2$.
\end{IEEEproof}

Unlike Shannon's coding theorem for AWGN channels 
  \cite[Theorem 9.1.1, p.~268]{cover02IT}, we cannot purge codewords from
  our codebook to lower the maximal decoding error probability, as
  that would violate the i.i.d.~condition for the codeword construction that 
  is needed to limit Willie's detection ability in our proof.
However, it is reasonable that users in sensitive situations attempting
  to hide their communications would prefer uniform rather than average
  decoding error performance, in essence demanding that the specific
  codebook they are using is effective.
In such a scenario, the construction of Theorem 
  \ref{th:achievability_peak_power} can be employed using the modification 
  given by the remark following its proof.
This construction also
  satisfies both the peak and the average power constraints, 
  as demonstrated below.
\begin{customtheorem}[Achievability under a peak power constraint]
\label{th:achievability_peak_power}
Suppose Alice's transmitter is subject to the peak power constraint $b$,
  $0<b<\infty$, and
  Willie's channel is subject to AWGN with power $\sigma^2_w>0$.
Also suppose that Alice and Bob share a secret of sufficient length.
Then Alice can maintain Willie's sum of the probabilities of 
  detection errors $\alpha+\beta\geq 1-\epsilon$ for any $\epsilon>0$
  while reliably transmitting $o(\sqrt{n})$ bits to Bob
  over $n$ uses of an AWGN channel when $\sigma_w^2$ is 
  unknown and $\mathcal{O}(\sqrt{n})$ bits in $n$ channel uses if she knows a
  lower bound $\sigma_w^2\geq\hat{\sigma}_w^2$ for some $\hat{\sigma}_w^2>0$.
\end{customtheorem}

To prove Theorem \ref{th:achievability_peak_power}, we introduce 
  a variant of the Leibniz integral rule as a lemma:
\begin{lemma}[Leibniz integral rule]\label{lemma:leibniz}
Suppose that $f(x,a)$ is defined for $x\geq x_0$ and $a\in [u,v],u<v$, and
  satisfies the following properties:
\begin{enumerate}
\item \label{cond:cont_integrand} $f(x,a)$ is continuous on $[u,v]$ for $x\geq x_0$;
\item \label{cond:cont_deriv} $\frac{\partial f(x,a)}{\partial a}$ is continuous on $[u,v]$ for $x\geq x_0$;
\item \label{cond:dom_integrand} There is a function $g(x)$  
  such that $|f(x,a)|\leq g(x)$ and $\int_{x_0}^{\infty} g(x)dx<\infty$;
\item \label{cond:dom_deriv} There is a function $h(x)$
  such that $|\frac{\partial f(x,a)}{\partial a}|\leq h(x)$ and
  $\int_{x_0}^{\infty} h(x)dx<\infty$.
\end{enumerate}
Then $\frac{\partial}{\partial a}\int_{x_0}^\infty f(x,a)dx=\int_{x_0}^\infty\frac{\partial f(x,a)}{\partial a}dx$.
\end{lemma}
The proof of Lemma \ref{lemma:leibniz} is available in 
  \cite[Ch.~XIII.3]{lang97ugradanalysis}.
We now prove Theorem \ref{th:achievability_peak_power}.

\begin{IEEEproof}[Proof (Theorem \ref{th:achievability_peak_power})]
\textbf{Construction:} 
Alice encodes the input in blocks of length $M$ bits into codewords of length $n$
  at the rate $R=M/n$ bits/symbol with the symbols drawn
  from alphabet $\{-a,a\}$, where $a$ satisfies the peak power constraint 
  $a^2<b$ and is defined later. %
We independently generate $2^{nR}$ codewords 
  $\{\mathbf{c}(W_k),k=1,2,\ldots,2^{nR}\}$ for messages
  $\{W_k\}$ from $\{-a,a\}^{n}$ according to 
  $p_{\mathbf{X}}(\mathbf{x})=\prod_{i=1}^{n}p_X(x_i)$, where
  $p_X(-a)=p_X(a)=\frac{1}{2}$.
As in the proof of Theorem \ref{th:achievability},
  this single-use codebook is not revealed to Willie, though he knows how it is constructed,
  including the value of $a$.
While the entire codebook is secretly shared between Alice and Bob,
  in the remark following the proof we discuss how to reduce the
  amount of shared secret information. 

\textbf{Analysis:}
When Alice transmits a symbol during the $i^{\text{th}}$ symbol period, 
  she transmits $-a$ or $a$ equiprobably by construction and
  Willie observes the symbol corrupted by AWGN.
Therefore,
$\mathbb{P}_s=\frac{1}{2}\left(\mathcal{N}(-a,\sigma_w^2)+\mathcal{N}(a,\sigma_w^2)\right)$,
  and, with $\mathbb{P}_w=\mathcal{N}(0,\sigma_w^2)$, we have:
\begin{align}
\label{eq:kl_peak_power}\mathcal{D}(\mathbb{P}_w\|\mathbb{P}_s)&=\int_{-\infty}^{\infty}\frac{e^{-\frac{x^2}{2\sigma_w^2}}}{\sqrt{2\pi}\sigma_w}\ln\frac{e^{-\frac{x^2}{2\sigma_w^2}}}{\frac{1}{2}\left(e^{-\frac{(x+a)^2}{2\sigma_w^2}}+e^{-\frac{(x-a)^2}{2\sigma_w^2}}\right)}dx
\end{align}
Since \eqref{eq:kl_peak_power} is an even function, we assume $a\geq 0$.

While there is no closed-form expression for 
  \eqref{eq:kl_peak_power}, its integrand is well-behaved, allowing 
  the application of Lemma \ref{lemma:taylor} to \eqref{eq:kl_peak_power}.
The Taylor series expansion with respect to $a$ around $a=0$ 
  can be performed using Lemma \ref{lemma:leibniz}.
We demonstrate that the conditions for Lemmas \ref{lemma:taylor} and 
  \ref{lemma:leibniz} hold in Appendix \ref{app:conds}.
The zeroth through third order terms of the Taylor series expansion of 
  \eqref{eq:kl_peak_power} are zero, as is the fifth term.
The fourth order term is:
\begin{eqnarray*}
\frac{a^4}{4!}\times\left.\frac{\partial^4 \mathcal{D}(\mathbb{P}_w\|\mathbb{P}_s)}{\partial a^4}\right|_{a=0}&=&\frac{a^4}{4\sigma_w^4}
\end{eqnarray*}
Suppose Alice sets $a^2\leq\frac{cf(n)}{\sqrt{n}}$, where $c$ and $f(n)$ are 
  defined as in Theorem \ref{th:achievability}.
The sixth derivative of \eqref{eq:kl_peak_power} with respect to $a$ is:
\begin{align}
\label{eq:D6_kl_peak_power}\frac{\partial^6 \mathcal{D}(\mathbb{P}_w\|\mathbb{P}_s)}{\partial a^6}&=-\int_{-\infty}^{\infty}\frac{8x^6e^{-\frac{x^2}{2\sigma_w^2}}}{\sigma_w^{12}\sqrt{2\pi}\sigma_w}\left(15\sech^6\left(\frac{ax}{\sigma_w^2}\right)-15\sech^4\left(\frac{ax}{\sigma_w^2}\right)+2\sech^2\left(\frac{ax}{\sigma_w^2}\right)\right)dx
\end{align}
where $\sech(x)=\frac{2}{e^x+e^{-x}}$ is the hyperbolic secant function.
Evaluated at zero, the sixth derivative is 
  $\left.\frac{\partial^6 \mathcal{D}(\mathbb{P}_w\|\mathbb{P}_s)}{\partial a^6}\right|_{a=0}=-\frac{240}{\sigma_w^6}$.
Since \eqref{eq:D6_kl_peak_power} is continuous (see Appendix \ref{app:conds}), 
  there exists a neighborhood $[0,\mu]$ such that, for all 
  $\xi\in[0,\mu]$, the remainder term
  $\frac{a^6}{6!}\times\left.\frac{\partial^6 \mathcal{D}(\mathbb{P}_w\|\mathbb{P}_s)}{\partial a^6}\right|_{a=\xi}\leq0$.
Then, for $n$ large enough, we can apply Lemma \ref{lemma:taylor} to 
  upper-bound relative entropy with the fourth order term as follows:
\begin{eqnarray}
\label{eq:tv_taylor_bound_pp}\mathcal{V}_T(\mathbb{P}_w^n,\mathbb{P}_s^n)\leq \frac{a^2}{2\sigma_w^2}\sqrt{\frac{n}{2}}\leq \frac{\epsilon f(n)}{\sigma_w^2}
\end{eqnarray}

Since the power of Alice's symbol is $a^2=P_f$,
  (\ref{eq:tv_taylor_bound_pp}) is identical to (\ref{eq:tv_taylor_bound})
  and Alice obtains the upper bound
  $\mathcal{V}_T(\mathbb{P}_w^n,\mathbb{P}_s^n)\leq \epsilon$,
  limiting the performance of Willie's detector.

Next let's examine the probability $\mathbb{P}_e$ of Bob's decoding 
  error averaged over all possible codebooks.
As in Theorem \ref{th:achievability}, we cannot directly apply the standard
  constant-power channel coding results to our system where the symbol power
  is a decreasing function of the codeword length.
We upper-bound Bob's decoding error probability
  by analyzing a suboptimal decoding scheme.
Suppose Bob uses a hard-decision device on each received 
  symbol $y^{(b)}_i=f_i+z^{(b)}_i$ via the rule 
  $\hat{f}_i=\left\{a\text{~if~}y^{(b)}_i\geq 0;-a\text{~otherwise}\right\}$,
  and applies an ML decoder on its output.
The effective channel for the encoder/decoder pair is a binary symmetric 
  channel with cross-over probability $p_e=Q(a/\sigma_b)$ and
  the probability of the decoding error averaged over all possible 
  codebooks is $\mathbb{P}_e\leq 2^{nR-n(1-\mathcal{H}(p_e))}$ 
  \cite{barg02randomcodeexponents},
  where $\mathcal{H}(p)=-p\log_2 p-(1-p)\log_2 (1-p)$ is the binary entropy
  function.
We expand the analysis in \cite[Section I.2.1]{majani88thesis} to characterize
  the rate $R$.
We use Lemma \ref{lemma:taylor} to upper-bound
  $p_e\leq\frac{1}{2}-\frac{1}{\sqrt{2\pi}}\left(\frac{a}{\sigma_b}-\frac{a^3}{6\sigma_b^3}\right)\triangleq p_e^{(UB)}$, 
  where $p_e^{(UB)}$ is the sum of the zeroth through second terms 
  of the Taylor series expansion of $Q(a/\sigma_b)$ around $a=0$.
The remainder term is non-positive for $a/\sigma_b$ satisfying
  $\frac{8a^6}{\sigma_b^6}-\frac{60a^4}{\sigma_b^4}+\frac{90a^2}{\sigma_b^2}-15\leq 0$,
  and, since $a^2=\frac{cf(n)}{\sqrt{n}}$, the upper bound thus holds for
  large enough $n$.
Since $\mathcal{H}(p)$ is a monotonically increasing function on the interval
  $\left[0,\frac{1}{2}\right]$, $\mathcal{H}(p_e)\leq\mathcal{H}(p_e^{(UB)})$.
The Taylor series expansion of $\mathcal{H}(p_e^{(UB)})$ with 
  respect to $a$ around $a=0$ yields 
  $\mathcal{H}(p_e^{(UB)})=1-\frac{a^2}{\sigma_b^2\pi\ln 2}+\mathcal{O}(a^4)$.
Substituting $a^2=\frac{cf(n)}{\sqrt{n}}$, we obtain 
  $\mathbb{P}_e\leq 2^{nR-\frac{\sqrt{n}cf(n)}{\sigma_b^2\pi\ln2}+\mathcal{O}(1)}$.
Since $f(n)=\omega(1/\sqrt{n})$,
  if rate $R=\frac{\rho cf(n)}{\sqrt{n}\sigma_b^2\pi\ln 2}$ bits/symbol for 
  a constant $\rho<1$,
  the probability of Bob's decoding error averaged over all codebooks
  decays exponentially to zero as
  $n$ increases and Bob obtains $nR=o(\sqrt{n})$ bits in $n$ channel
  uses, and $\mathcal{O}(\sqrt{n})$ bits in $n$ channel uses if $f(n)=\hat{\sigma}_w^2$.
\end{IEEEproof}

\subsection*{Remarks}

\subsubsection*{Employing the best codebook}
The proof of Theorem \ref{th:achievability_peak_power} guarantees Bob's 
  decoding error performance averaged over all binary codebooks.
Following the standard coding arguments \cite[p.~204]{cover02IT}, 
  there must be at least one binary alphabet codebook that has at least 
  average probability of error.
Thus, to guarantee uniform performance, Alice and Bob must select
  ``good'' codebooks for communications.
However, choosing specific codebooks
  would violate the i.i.d.~condition for the codeword construction
  that is needed to limit Willie's detection capability in our proof.

Consider a codebook that has at least
  average probability of error, but now assume that it is public 
  (i.e.~known to Willie).
Theorem \ref{th:achievability_peak_power} shows that Alice can use it to
  transmit $\mathcal{O}(\sqrt{n})$ bits to Bob in $n$ channel uses with
  exponentially-decaying probability of error.
However, since the codebook is public, unless Alice and Bob take steps to
  protect their communication, Willie can use this codebook to detect Alice's
  transmissions by performing the same decoding as Bob.
Here we demonstrate that to use a public codebook it suffices for Alice and Bob
  to share a secret random binary vector and note that this resembles the
  one-time pad scheme from traditional cryptography \cite{shannon49sec},
  but employed here for a very different application.

Suppose that, prior to communication, Alice and Bob generate and share 
  binary vector $\mathbf{k}$ where
  $p_{\mathbf{K}}(\mathbf{k})=\prod_{i=1}^{n}p_K(k_i)$ with
  $p_K(0)=p_K(1)=\frac{1}{2}$.
Alice XORs $\mathbf{k}$ and the binary representation of the codeword 
  $\mathbf{c}(W_k)$, resulting in an equiprobable transmission of $-a$ and $a$
  when Alice transmits a symbol during the $i^{\text{th}}$ symbol
  period.
Provided $\mathbf{k}$ is never re-used and is kept secret from Willie, 
  the i.i.d.~assumption for 
  the vector $\mathbf{y}_w$ in Theorem \ref{th:achievability_peak_power} 
  holds %
  without the need to exchange an entire secret codebook between Alice and Bob.
Bob decodes by XORing $\mathbf{k}$ with the output of the hard-decision device
  prior to applying the ML decoder.
While the square root law implies that the shared $\mathcal{O}(n)$-bit secret 
  here is quadratic in 
  the length $M=\mathcal{O}(\sqrt{n})$ of a message, we offer a coding scheme 
  that, on average, requires an
  $\mathcal{O}(\sqrt{n}\log n)$-bit secret in Appendix \ref{app:key}.
The development of LPD communication with a shared secret either linear
  or sublinear in the message size is a subject of future research.

\subsubsection*{Relationship with Square Root Law in Steganography}
The LPD communication problem is related to the problem of steganography.
A comprehensive review of steganography is available in a book by Fridrich 
  \cite{fridrich09stego}.
In finite-alphabet imperfect steganographic systems 
  at most $\mathcal{O}(\sqrt{n})$
  symbols in the original covertext of length $n$ may safely be modified to hide
  a steganographic message of length $\mathcal{O}(\sqrt{n}\log{n})$ bits
  \cite[Ch.~13]{fridrich09stego} \cite{ker07sqrtlaw}.
This result was extended to Markov covertext \cite{filler09sqrtlawmarkov} and
  was shown to either require a key linear in the size
  of the message \cite{ker09sqrtlawkey} or encryption of the message prior to 
  embedding \cite{ker10sqrtlawnokey}.

The square root law in steganography 
  has the same form as our square root law
  because both laws follow from %
  the property that relative entropy is locally quadratic 
  \cite[Ch.~2.6]{kullback59IT}: 
\begin{eqnarray*}
\mathcal{D}(\mathbb{P}_{0}\|\mathbb{P}_{1})&=&\frac{\delta^2}{2}\mathcal{J}(\theta)+\mathcal{O}(\delta^3)
\end{eqnarray*}
where 
  $\mathcal{J}(\theta)=\int_{\mathcal{X}}\left(\frac{\partial}{\partial\theta}\ln f(x;\theta)\right)^2f(x;\theta)dx$
  is the Fisher information associated with parameter $\theta$, and
  $\mathbb{P}_{0}$ and $\mathbb{P}_{1}$ are probability measures with
  density functions from the same family over the support $\mathcal{X}$, 
  but with parameters differing by $\delta$: $p_0(x)=f(x;\theta)$ and 
  $p_1(x)=f(x;\theta+\delta)$.
Fisher information is thus used as a metric for 
  steganographic security \cite{filler09fisherstego,ker09fisherinfoest}.
 
In a typical steganography scenario with a passive warden,
  coding techniques similar to Hamming codes allow
  embedding of $\log(n)$ bits per changed symbol 
  \cite[Ch.~8]{fridrich09stego}, which make hiding
  $\mathcal{O}(\sqrt{n}\log{n})$ bits in $n$ symbols possible.
However, due to the noise on the channel between Alice and Bob,
  and the resultant need for error correction,
  the LPD channel only allows $\mathcal{O}(\sqrt{n})$ bits to 
  be transmitted in $n$ channel uses, as we prove in the following section.

\section{Converse}
\label{sec:converse}
Here, as in the proof of achievability, the channel between Alice and 
  Bob is AWGN with power $\sigma_b^2$.
Alice's objective is to transmit a message $W_k$ that is
  $M=\omega(\sqrt{n})$ bits long
  to Bob in $n$ channel uses with arbitrarily small probability of decoding 
  error as $n$ gets large, while limiting Willie's ability to detect
  her transmission.
Alice encodes each message $W_k$ arbitrarily into $n$ symbols at the rate 
  $R=M/n$ symbols/bit.
For an upper bound on the reduction in entropy, the messages are chosen
  equiprobably.

Willie observes all $n$ of Alice's channel uses, but
  he is oblivious to her signal properties and employs only a simple
  power detector.
Nevertheless, we prove that, even if Willie only has these limited capabilities,
  Alice cannot transmit a 
  message with $\omega(\sqrt{n})$ bits of information
  in $n$ channel uses without either being detected by Willie or having Bob 
  suffer a non-zero decoding error probability.

\setcounter{theorem}{1}
\begin{theorem}
If over $n$ channel uses, Alice attempts to transmit a message to Bob
  that is $\omega(\sqrt{n})$ bits long, 
  then, as $n\rightarrow\infty$, either there exists a detector that Willie 
  can use to detect her 
  with arbitrarily low sum of error probabilities $\alpha+\beta$,
  or Bob cannot decode with arbitrarily low
  probability of error.
\end{theorem}

\begin{IEEEproof}
Suppose Alice employs an arbitrary codebook 
  $\{\mathbf{c}(W_k), k=1,2,\ldots,2^{nR}\}$.
Detection of Alice's transmissions entails Willie deciding between the 
  following hypotheses:
\begin{eqnarray*}
H_0:& &y^{(w)}_i=z^{(w)}_i,~i=1,\ldots,n\\
H_1:& &y^{(w)}_i=f_i+z^{(w)}_i,~i=1,\ldots,n
\end{eqnarray*}
Suppose Willie uses a power detector to perform the hypothesis test as follows:
  first, he collects a row vector of $n$ independent readings $\mathbf{y}_w$ 
  from his channel to Alice.
Then he generates the test statistic 
  $S=\frac{\mathbf{y}_w\mathbf{y}_w^T}{n}$
  where $\mathbf{x}^T$ denotes the transpose of vector $\mathbf{x}$,
  and rejects or accepts the null hypothesis based on a comparison of $S$
  to a threshold that we discuss later.
We first show how Willie can bound the error probabilities $\alpha$ and 
  $\beta$ of the power detector as a function of Alice's signal parameters.
Then we show that if Alice's codebook prevents Willie's test from detecting her,
  Bob cannot decode her transmissions without error.

If the null hypothesis $H_0$ is true, Alice does not transmit and
  Willie observes AWGN on his channel.
Thus, $y^{(w)}_i\sim\mathcal{N}(0,\sigma_w^2)$, and
  the mean and the variance of $S$ when $H_0$ is
  true are:
\begin{eqnarray}
\label{eq:P_0_mean}\expected{S}&=&\sigma_w^2\\
\label{eq:P_0_var}\Var\left[S\right]&=&\frac{2\sigma_w^4}{n}
\end{eqnarray}

Suppose Alice transmits codeword 
  $\mathbf{c}(W_k)=\{f_i^{(k)}\}_{i=1}^n$.
Then Willie's vector of observations
  $\mathbf{y}_{w,k}=\{y^{(w,k)}_i\}_{i=1}^n$ contains readings of 
  mean-shifted noise 
  $y^{(w,k)}_i\sim\mathcal{N}(f_i^{(k)},\sigma_w^2)$.
The mean of each squared observation is 
$\expected{y_i^2}=\sigma_w^2+\left(f_i^{(k)}\right)^2$ and the variance is 
$\Var\left[y_i^2\right]=\expected{y_i^4}-\left(\expected{y_i^2}\right)^2=4\left(f_i^{(k)}\right)^2\sigma_w^2+2\sigma_w^4$.
Denote the average symbol power of codeword $\mathbf{c}(W_k)$ by
$P_k=\frac{\mathbf{c}(W_k)\mathbf{c}^T(W_k)}{n}$.
Then the mean and variance of $S$ when Alice
  transmits codeword $\mathbf{c}(W_k)$ are:
\begin{eqnarray}
\label{eq:P_1_mean}\expected{S}&=&\sigma_w^2+P_k\\
\label{eq:P_1_var}\Var\left[S\right]&=&\frac{4P_k\sigma_w^2+2\sigma_w^4}{n}
\end{eqnarray}
The variance of Willie's test statistic (\ref{eq:P_1_var}) is computed by
  adding the variances conditioned on $\mathbf{c}(W_k)$ of the squared 
  individual observations $\Var\left[y_i^2\right]$ 
  (and dividing by $n^2$) since the noise on the individual observations
  is independent.

The probability distribution for the vector of 
  Willie's observations depends on which hypothesis is true.
Denote by $\mathbb{P}_0$ the distribution when $H_0$ holds, and 
  $\mathbb{P}_1^{(k)}$ when 
  $H_1$ holds with Alice transmitting message $W_k$.
While $\mathbb{P}_1^{(k)}$ is conditioned on Alice's codeword, 
  we show that the average symbol power 
  $P_k=\frac{\mathbf{c}(W_k)\mathbf{c}^T(W_k)}{n}$
  of codeword $\mathbf{c}(W_k)$ determines its detectability by this detector,
  and that our result applies to all codewords with power of the same order.

If $H_0$ is true, then $S$
  should be close to (\ref{eq:P_0_mean}).
Willie picks a threshold $t$ and 
  compares the value of $S$ to $\sigma_w^2+t$.
He accepts $H_0$ if $S< \sigma_w^2+t$ and rejects it otherwise.
Suppose that he desires false positive probability $\alpha^*$,
  which is the probability that $S\geq \sigma_w^2+t$ when $H_0$
  is true.
We bound it using (\ref{eq:P_0_mean}) and (\ref{eq:P_0_var}) 
  with Chebyshev's Inequality \cite[Eq.~(3.32)]{cover02IT}:
\begin{eqnarray*}
\alpha &=&\mathbb{P}_0\left(S\geq \sigma_w^2+t\right)\\
&\leq&\mathbb{P}_0\left(|S- \sigma_w^2|\geq t\right)\\
&\leq&\frac{2\sigma_w^4}{nt^2}
\end{eqnarray*}
Thus, to obtain $\alpha^*$, Willie sets $t=\frac{d}{\sqrt{n}}$, where 
  $d=\frac{\sqrt{2}\sigma_w^2}{\sqrt{\alpha^*}}$ is a 
  constant.
As $n$ increases, $t$ decreases, which
  is consistent with Willie gaining greater confidence with more observations.

Suppose Alice transmits codeword $\mathbf{c}(W_k)$.
Then the probability of a miss $\beta^{(k)}$
  is the probability that $S< \sigma_w^2+t$, where $t=\frac{d}{\sqrt{n}}$.
We bound $\beta^{(k)}$ using (\ref{eq:P_1_mean}) 
  and (\ref{eq:P_1_var}) with Chebyshev's Inequality:
\begin{eqnarray}
\beta^{(k)}&=&\mathbb{P}_1^{(k)}\left(S< \sigma_w^2+t\right)\nonumber\\
&\leq&\mathbb{P}_1^{(k)}\left(\left|S-\sigma_w^2-P_k\right|\geq P_k-t\right)\nonumber\\
\label{eq:beta_bound}&\leq&\frac{4P_k\sigma_w^2+2\sigma_w^4}{(\sqrt{n}P_k-d)^2}
\end{eqnarray}
If the average symbol power $P_k=\omega(1/\sqrt{n})$, 
  $\lim_{n\rightarrow\infty}\beta^{(k)}=0$.
Thus, with enough observations, Willie can detect 
  with arbitrarily low error probability Alice's codewords with 
  the average symbol power 
  $P_k=\frac{\mathbf{c}(W_k)\mathbf{c}^T(W_k)}{n}=\omega(1/\sqrt{n})$.
Note that Willie's detector is oblivious to any details of Alice's codebook 
  construction.

On the other hand, if the transmitted codeword has the average symbol power 
  $P_{\mathcal{U}}=\mathcal{O}(1/\sqrt{n})$, then
  \eqref{eq:beta_bound} does not upper-bound the probability of a 
  missed detection arbitrarily close to zero regardless of the number of 
  observations.
Thus, if Alice desires to lower-bound the sum of the 
  probabilities of error of Willie's statistical test by 
  $\alpha+\beta\geq\zeta>0$, her codebook must contain a positive fraction 
  $\gamma$ of such low-power codewords.
Let's denote this subset of codewords with the average symbol power 
  $P_{\mathcal{U}}=\mathcal{O}(1/\sqrt{n})$ as $\mathcal{U}$
  and examine the probability of Bob's decoding error $\mathbb{P}_e$.
The probability that a message from set $\mathcal{U}$ is sent is 
  $\mathbb{P}\left(\mathcal{U}\right)=\gamma$, as all messages are equiprobable.
We bound %
$\mathbb{P}_e=\mathbb{P}_e\left(\mathcal{U}\right)\prob{\mathcal{U}}+\mathbb{P}_e\left(\overline{\mathcal{U}}\right)\prob{\overline{\mathcal{U}}}\geq\gamma\mathbb{P}_e\left(\mathcal{U}\right)$,
where $\overline{\mathcal{U}}$ is the complement of $\mathcal{U}$ and 
  $\mathbb{P}_e\left(\mathcal{U}\right)$ is the probability of decoding 
  error when a message from $\mathcal{U}$ is sent:
\begin{eqnarray}
\label{eq:temp1}\mathbb{P}_e\left(\mathcal{U}\right)&=&\frac{1}{|\mathcal{U}|}\sum_{W\in \mathcal{U}}\mathbb{P}_e\left(\mathbf{c}(W)\text{~sent}\right)
\end{eqnarray}
where $\mathbb{P}_e\left(\mathbf{c}(W)\text{~sent}\right)$ is the 
  probability of error when codeword $\mathbf{c}(W)$ is transmitted,
  $|\cdot|$ denotes the set cardinality operator, and (\ref{eq:temp1}) holds
  because all messages are equiprobable.

When Bob uses the optimal decoder, 
  $\mathbb{P}_e\left(\mathbf{c}(W)\text{~sent}\right)$ 
  is the probability that Bob decodes the received signal as $\hat{W}\neq W$.
This is the probability of a union
  of events $E_j$, where $E_j$ is the event that sent message $W$ is 
  decoded as some other message $W_j\neq W$:
\begin{eqnarray}
\mathbb{P}_e\left(\mathbf{c}(W)\text{~sent}\right)&=&\prob{\cup_{j=1,W_j\neq W}^{2^{nR}}E_j}\nonumber\\
\label{eq:decoding_error}&\geq&\prob{\cup_{W_j\in\mathcal{U}\backslash\{W\}}E_j}\triangleq \mathbb{P}^{(\mathcal{U})}_e
\end{eqnarray}
Here the inequality in (\ref{eq:decoding_error}) is due to the observation that
  the sets in the second union are contained in the first. 
From the decoder perspective, this is due to the decrease in the decoding 
  error probability
  if Bob knew that the message came from $\mathcal{U}$ (reducing
  the set of messages on which the decoder can err).

Our analysis of $\mathbb{P}^{(\mathcal{U})}_e$ uses Cover's simplification of 
  Fano's inequality similar to the proof of
  the converse to the coding theorem for Gaussian channels 
  in \cite[Ch.~9.2]{cover02IT}.
Since we are interested in $\mathbb{P}^{(\mathcal{U})}_e$, we do not absorb
 it into $\epsilon_n$ as done in (9.37) of \cite{cover02IT}.
 Rather, we explicitly use:
\begin{eqnarray}
\label{eq:cover_fano}H(W|\hat{W})&\leq&1+(\log_2|\mathcal{U}|)\mathbb{P}^{(\mathcal{U})}_e
\end{eqnarray}
where $H(W|\hat{W})$ denotes the entropy of message $W$ conditioned on Bob's
  decoding $\hat{W}$ of $W$.

Noting that the size of the set $\mathcal{U}$ from which the messages
  are drawn is $\gamma 2^{n R}$ and that, since each message is equiprobable, 
  the entropy of a message $W$ from 
  $\mathcal{U}$ is $H(W)=\log_2|\mathcal{U}|=\log_2\gamma + nR$,
  we utilize (\ref{eq:cover_fano}) and carry out steps (9.38)--(9.53) in
  \cite{cover02IT} to obtain:
\begin{eqnarray}
\label{eq:P_e_U}\mathbb{P}^{(\mathcal{U})}_e&\geq&1-\frac{P_{\mathcal{U}}/2\sigma_b^2+1/n}{\frac{\log_2\gamma}{n}+R}
\end{eqnarray}
Since Alice transmits $\omega(\sqrt{n})$ bits in $n$ channel
  uses, her rate is $R=\omega(1/\sqrt{n})$ bits/symbol.
However, $P_{\mathcal{U}}=O(1/\sqrt{n})$, and,
  as $n\rightarrow \infty$, %
  $\mathbb{P}^{(\mathcal{U})}_e$ is bounded away from zero.
Since $\gamma>0$,
  $\mathbb{P}_e$ is bounded away from zero if Alice tries to transmit 
  $\omega(\sqrt{n})$ bits reliably while beating Willie's
  simple power detector.
\end{IEEEproof}

\subsection*{Goodput of Alice's Communication}
Define the goodput $G(n)$ of Alice's communication as the average
  number of bits that Bob can receive from Alice 
  over $n$ channel uses with non-zero probability of a message being 
  undetected as $n\rightarrow\infty$.
Since only $\mathcal{U}$ contains such messages, by (\ref{eq:P_e_U}),
  the probability of her message being successfully decoded by 
  Bob is
  $\mathbb{P}^{(\mathcal{U})}_s=1-\mathbb{P}^{(\mathcal{U})}_e=\mathcal{O}\left(\frac{1}{\sqrt{n}R}\right)$
  and the goodput is $G(n)=\gamma\mathbb{P}^{(\mathcal{U})}_sRn=\mathcal{O}(\sqrt{n})$.
Thus, Alice cannot break the square root law using an arbitrarily high 
  transmission rate and retransmissions while keeping the power below
  Willie's detection threshold.

\section{Discussion}
\label{sec:discussion}

\subsection{Relationship to Previous Work in Communications}
The relationship of our work to steganography has already been discussed in 
  the remark at the end of Section \ref{sec:achievability}.
Here we relate our problem to other work in communication.

\subsubsection*{Spread Spectrum Communications}
As wireless communication became
  prevalent, militaries sought methods to protect their signals from being 
  detected by the enemy, leading to
  the development of spread spectrum communication.
Spread spectrum communication provides an LPD capability as well as resistance
  to jamming by transmitting a signal that 
  requires bandwidth $W_M$ on a much wider bandwidth 
  $W_s\gg W_M$, thereby reducing the power 
  spectral density.  
Most spread spectrum results address the practical aspects of spread spectrum 
  architectures and comprehensive reviews \cite{simon94ssh,varakin85ss} 
  are available. 
We are not aware of any prior work studying the fundamental limits on 
  the information that can be transmitted with low probability of detection
  using spread spectrum technology.
However, we note that, while we present our result for narrowband channels, 
  our analysis trivially translates to wideband channels as well: Alice can 
  reliably transmit 
  $\mathcal{O}(\sqrt{W_s n})$ LPD bits per $n$ uses of a channel with 
  bandwidth $W_s$.
Thus, spread spectrum systems are also limited by the square root law.

\subsubsection*{Information-theoretic secrecy}
There exists a rich body of literature on the information-theoretic secrecy 
  resulting from the legitimate receiver having a better channel to
  the transmitter than the adversary.
Wyner was the first to show that if the adversary only has access to a noisy 
  version
  of the signal received by the legitimate receiver (using a \emph{wire-tap 
  channel}), then the legitimate receiver can achieve a positive secure 
  communication rate to the sender without the use of a shared one-time pad 
  \cite{wyner75wiretap}.
Cheong and Hellman extended this result to Gaussian channels 
  \cite{leung-yan-cheong78gaussianwiretap}, and Csisz\'{a}r and K\"{o}rner
  generalized it to broadcast channels \cite{csiszar78broadcastwiretap}.
Our approach considers the adversary's ability to detect rather than decode the
  transmissions, and it does not rely on the channel to the legitimate
  receiver being better than the channel to the adversary.
However, recent succeeding work \cite{che13sqrtlawbsc} claims that if 
  the adversary and the legitimate
  receiver each has a \emph{binary symmetric channel} (BSC) to the transmitter,
  with the adversary having a significantly noisier channel (i.e.~a wire-tap 
  BSC with positive secrecy rate), then the square-root law of LPD 
  communication is achievable without the use of a secret codebook.

\subsubsection*{Anonymous communication}
Our problem is related to that of anonymous communication 
  \cite{danezis08anonymouscommsurvey}, specifically the task of defeating
  the network traffic timing analysis.
While the objective is fundamentally the same, the setting and approaches are 
  vastly different.
The network traffic analysis involves the adversary inferring network properties
  (such as source-relay pairs) by correlating properties (such as the 
  inter-packet timing) of two or more encrypted 
  packet flows.
Protecting against this kind of analysis is costly, as one needs to make
  flows look statistically independent by
  randomizing the timing of the packets, inserting dummy packets, 
  or dropping a portion of the data packets.
Recent work thus addressed the amount of common information that can be
  embedded into two flows that are generated by independent renewal processes
  \cite{marano12renewaltraffic}.
However, in our scenario Willie cannot perform traffic analysis
  (or any kind of network layer analysis), 
  as Alice prevents him (with high probability)
  from detecting her transmission in the first place.

\subsubsection*{Cognitive Radio}
The LPD communication problem is also related to that of 
  establishing a cognitive radio (CR) network \cite{yucek09crsensingsurvey}.
An aspect of the CR problem is limiting the interference from the secondary 
  users' radios to the primary users of the network.
The LPD problem with a passive warden can be cast within this framework by 
  having primary users only listen \cite{ren11coexistence}.
However, the properties of the secondary signal that allow smooth operation of 
  the primary network are very different from those of an undetectable signal.
While there is a lot of work on the former topic, we are not aware of work 
  by the CR community on the latter issue.

\subsection{Impact of Adversary's \emph{a priori} Knowledge of the Transmission State on Achievability}
\label{sec:bayes}

The proofs of achievability (Theorems \ref{th:achievability} and 
  \ref{th:achievability_peak_power}) in Section \ref{sec:achievability}
  assume that Willie has no prior knowledge on
  whether Alice transmits or not.
Here we argue that the assumption of a non-trivial prior distribution on Alice's
  transmission state does not impact our asymptotic results.
Suppose that Willie knows that Alice does not transmit
  (i.e.~$H_0$ is true) with probability $\pi_0$ and that she transmits
  (i.e.~$H_0$ is true) with probability $\pi_1=1-\pi_0$.
Let $\mathbb{P}_e$ denote the probability that Willie's hypothesis test makes
  an error averaged over all observations.
The following generalized version of Fact \ref{fact:totvar_testperf} then holds:
\begin{fact}[Generalized Fact \ref{fact:totvar_testperf}]  
\label{fact:totvar_testperf_bayes}
$\mathbb{P}_e\geq\min(\pi_0,\pi_1)-\max(\pi_0,\pi_1)\mathcal{V}_T(\mathbb{P}_0,\mathbb{P}_1)$
\end{fact}
\noindent where, as in Section \ref{sec:achievability}, we denote the 
  probability distribution of Willie's channel 
  observations conditioned on Alice not transmitting (i.e.~on $H_0$ being true)
  as $\mathbb{P}_0$, and the probability distribution of the 
  observations conditioned on Alice transmitting (i.e.~on $H_1$ being true) as 
  $\mathbb{P}_1$.
\ifx\techreport\undefined
The proof is available in \cite{bash12trsquarerootnewtitle}.
\else
The proof is in Appendix \ref{app:bayes}.
\fi
Thus, while Fact \ref{fact:totvar_testperf_bayes} demonstrates that additional 
  information about the
  likelihood of Alice transmitting helps Willie, the square root law still
  holds via the bounds
  on the total variation distance $\mathcal{V}_T(\mathbb{P}_0,\mathbb{P}_1)$.

\subsection{Mapping to a Continuous-time Channel}
\label{sec:ct_channel}
We employ a discrete-time model throughout 
  the paper.  
However, while this is commonly assumed without loss of generality 
  in standard communication theory, it is important to consider whether some 
  aspect of the LPD problem has been missed by focusing on discrete time.

Consider the standard communication system model, where Alice's (baseband) 
  continuous-time waveform is given in terms of her discrete time 
  transmitted sequence by:
\begin{eqnarray*}
x(t) = \sum_{i=1}^{n} f_i~p(t - i T_s)
\end{eqnarray*}
where $T_s$ is the symbol period and $p(\cdot)$ is the
pulse shaping waveform.  
Consider a (baseband) system bandwidth constraint of $W$ Hz.  
Now, if Alice chooses $p(\cdot)$ ideally as $\sinc(2 W t)$, where 
  $\sinc(x) =\frac{\sin(\pi x)}{\pi x}$, then the natural choice of
  $T_s = 1/2W$ results in no intersymbol interference (ISI).
From the Nyquist sampling criterion, both Willie (and Bob) can extract 
  all of the information from the signaling band by sampling at a rate of $2W$
  samples/second, which then leads directly to the discrete-time model 
  of Section \ref{sec:prerequisites} and suits our demonstration of the 
  fundamental limits to Alice's LPD channel capabilities.
However, when $p(\cdot)$ is chosen in a more practical fashion, for example, 
  as a raised cosine pulse with some excess bandwidth, then sampling at a rate 
  higher than $2W$ has utility for signal detection even if the Nyquist ISI 
  criterion is satisfied.  
In particular, techniques involving cyclostationary detection are now 
  applicable, and we consider such a scenario a promising area for future work.

\section{Conclusion}
\label{sec:conclusion}
Practitioners have always known that LPD communication
  requires one to use low power in order to blend in with the noise on 
  the eavesdropping warden's channel.
However, the specific requirements for achieving LPD communication and
  resulting achievable performance have not
  been analyzed prior to this work.
We quantified the conditions for existence and maintenance of an LPD channel
  by proving that the LPD communication is subject to 
  a square root law in that the number of LPD bits that can be 
  transmitted in $n$ channel uses is $\mathcal{O}(\sqrt{n})$.

There are a number of avenues for future research.
The key efficiency and, specifically, LPD communication with a secret linear 
  in the message length is an open theoretical research problem.
Practical network settings and the implications of the square root law on 
  the LPD transmission of packets under additional constraints such as delay
  should be analyzed.
The impact of dynamism in the network should also be examined, as
  well as more realistic scenarios that include channel artifacts such
  as fading and interference from other nodes.
One may be able to improve LPD communication by employing
  nodes that perform friendly jamming.
Eventually, we would like to answer this fundamental question: is it possible
  to establish and maintain a ``shadow'' wireless network in the presence of
  both active and passive wardens?

\appendix

\subsection{$\mathcal{D}(\mathbb{P}_w\|\mathbb{P}_s)$ in the proof of Theorem 
\ref{th:achievability_peak_power} meets the conditions of Lemmas 
\ref{lemma:taylor} and \ref{lemma:leibniz}}
\label{app:conds}

Re-arranging the terms of \eqref{eq:kl_peak_power} results in the following 
  expression:
\begin{align}
\label{eq:kl_peak_power_v2}\mathcal{D}(\mathbb{P}_w\|\mathbb{P}_s)&=\frac{a^2}{2\sigma_w^2}-\int_{-\infty}^{\infty}\frac{e^{-\frac{x^2}{2\sigma_w^2}}}{\sqrt{2\pi}\sigma_w}\ln\cosh\left(\frac{ax}{\sigma_w^2}\right)dx
\end{align}
where $\cosh(x)=\frac{e^{x}+e^{-x}}{2}$ is the hyperbolic cosine function.
Since $\frac{a^2}{2\sigma_w^2}$ is clearly continuous and differentiable
  with respect to $a$,
  we focus on the integral in \eqref{eq:kl_peak_power_v2}, specifically on
  its integrand:
\begin{align}
K(x,a)&=\frac{e^{-\frac{x^2}{2\sigma_w^2}}}{\sqrt{2\pi}\sigma_w}\ln\cosh\left(\frac{ax}{\sigma_w^2}\right)
\end{align}
Due to the peak power constraint, $0\leq a\leq \sqrt{b}$.
Also, $\ln\cosh(x)\leq |x|$ since 
  $\ln\left(\frac{e^x+e^{-x}}{2}\right)-|x|=\ln\left(\frac{1+e^{-2|x|}}{2}\right)\leq 0$.
Therefore, 
  $g(x)=\frac{\sqrt{b}|x|e^{-\frac{x^2}{2\sigma_w^2}}}{\sqrt{2\pi}\sigma_w^3}
  \geq |K(x,a)|$, in other words, $g(x)$ dominates $K(x,a)$.
$g(x)$ is integrable since
  $\int_{-\infty}^{\infty}g(x)dx=\sqrt{\frac{2b}{\pi\sigma_w^2}}<\infty$.

The derivatives of $K(x,a)$ with respect to $a$ can be written in the following
  form:
\begin{align}
\label{eq:dKodd}\text{odd~}i~:&\frac{\partial^i K(x,a)}{\partial a^i}=\frac{e^{-\frac{x^2}{2\sigma_w^2}}}{\sqrt{2\pi}\sigma_w}\frac{x^i}{\sigma_w^{2i}}\tanh\left(\frac{ax}{\sigma_w^2}\right)\sum_{k=1}^{(i-1)/2}c_{i,k}\sech^{2k}\left(\frac{ax}{\sigma_w^2}\right)\\
\label{eq:dKeven}\text{even~}i:~&\frac{\partial^i K(x,a)}{\partial a^i}=\frac{e^{-\frac{x^2}{2\sigma_w^2}}}{\sqrt{2\pi}\sigma_w}\frac{x^i}{\sigma_w^{2i}}\sum_{k=1}^{i/2}c_{i,k}\sech^{2k}\left(\frac{ax}{\sigma_w^2}\right)
\end{align}
where $\sech(x)=\frac{2}{e^{x}+e^{-x}}$ and 
  $\tanh(x)=\frac{e^x-e^{-x}}{e^{x}+e^{-x}}$ are the hyperbolic secant and 
  tangent functions, respectively, $c_{i,k}$ are constants, and the ``empty''
  sum $\sum_{k=1}^0 c_{i,k}=1$.
\ifx\techreport\undefined
We provide the first six derivatives in \cite{bash12trsquarerootnewtitle}.
\else
The first six derivatives of $K(x,a)$ with respect to $a$ are as follows:
\begin{align}
\label{eq:dK}\frac{\partial K(x,a)}{\partial a}&=\frac{e^{-\frac{x^2}{2\sigma_w^2}}}{\sqrt{2\pi}\sigma_w}\frac{x}{\sigma_w^2}\tanh\left(\frac{ax}{\sigma_w^2}\right)\\
\frac{\partial^2 K(x,a)}{\partial a^2}&=\frac{e^{-\frac{x^2}{2\sigma_w^2}}}{\sqrt{2\pi}\sigma_w}\frac{x^2}{\sigma_w^4}\sech^2\left(\frac{ax}{\sigma_w^2}\right)\\
\frac{\partial^3 K(x,a)}{\partial a^3}&=-\frac{e^{-\frac{x^2}{2\sigma_w^2}}}{\sqrt{2\pi}\sigma_w}\frac{2x^3}{\sigma_w^6}\sech^2\left(\frac{ax}{\sigma_w^2}\right)\tanh\left(\frac{ax}{\sigma_w^2}\right)\\
\frac{\partial^4 K(x,a)}{\partial a^4}&=\frac{e^{-\frac{x^2}{2\sigma_w^2}}}{\sqrt{2\pi}\sigma_w}\frac{2x^4}{\sigma_w^8}\left(2\sech^2\left(\frac{ax}{\sigma_w^2}\right)-3\sech^4\left(\frac{ax}{\sigma_w^2}\right)\right)\\
\frac{\partial^5 K(x,a)}{\partial a^5}&=\frac{e^{-\frac{x^2}{2\sigma_w^2}}}{\sqrt{2\pi}\sigma_w}\frac{8x^5\tanh\left(\frac{ax}{\sigma_w^2}\right)}{\sigma_w^{10}}\left(3\sech^4\left(\frac{ax}{\sigma_w^2}\right)-\sech^2\left(\frac{ax}{\sigma_w^2}\right)\right)\\
\label{eq:d6K}\frac{\partial^6 K(x,a)}{\partial a^6}&=\frac{e^{-\frac{x^2}{2\sigma_w^2}}}{\sqrt{2\pi}\sigma_w}\frac{8x^6}{\sigma_w^{12}}\left(15\sech^6\left(\frac{ax}{\sigma_w^2}\right)-15\sech^4\left(\frac{ax}{\sigma_w^2}\right)+2\sech^2\left(\frac{ax}{\sigma_w^2}\right)\right)
\end{align}
\fi
Clearly, $K(x,a)$ and its derivatives are
  continuous, satisfying conditions \ref{cond:cont_integrand} and 
  \ref{cond:cont_deriv} of Lemma \ref{lemma:leibniz}.
Since $-1\leq \tanh(x)\leq 1$ and $0\leq \sech(x)\leq 1$ for all
  real $x$, we can use the triangle inequality to show that 
  $\left|\frac{\partial^i K(x,a)}{\partial a^i}\right|\leq h_i(x)$
  where
\begin{align}
h_i(x)&=\frac{e^{-\frac{x^2}{2\sigma_w^2}}}{\sqrt{2\pi}\sigma_w}\frac{|x|^i}{\sigma_w^{2i}}\sum_{k=1}^{\lfloor i/2\rfloor}|c_{i,k}|
\end{align}
with $\lfloor x \rfloor$ denoting the largest integer $y\leq x$.
\ifx\techreport\undefined
We provide $h_i(x)$ for $i=1,\ldots,6$ in \cite{bash12trsquarerootnewtitle}.
\else
Therefore, the following relations show dominating functions of the
  corresponding derivatives of $K(x,a)$:
\begin{align}
\left|\frac{\partial K(x,a)}{\partial a}\right|&\leq h_1(x)=\frac{e^{-\frac{x^2}{2\sigma_w^2}}}{\sqrt{2\pi}\sigma_w}\frac{|x|}{\sigma_w^2}\\
\left|\frac{\partial^2 K(x,a)}{\partial a^2}\right|&\leq h_2(x)=\frac{e^{-\frac{x^2}{2\sigma_w^2}}}{\sqrt{2\pi}\sigma_w}\frac{|x|^2}{\sigma_w^4}\\
\left|\frac{\partial^3 K(x,a)}{\partial a^3}\right|&\leq h_3(x)=\frac{e^{-\frac{x^2}{2\sigma_w^2}}}{\sqrt{2\pi}\sigma_w}\frac{2|x|^3}{\sigma_w^6}\\
\label{eq:h4}\left|\frac{\partial^4 K(x,a)}{\partial a^4}\right|&\leq h_4(x)=\frac{e^{-\frac{x^2}{2\sigma_w^2}}}{\sqrt{2\pi}\sigma_w}\frac{10|x|^4}{\sigma_w^8}\\
\left|\frac{\partial^5 K(x,a)}{\partial a^5}\right|&\leq h_5(x)=\frac{e^{-\frac{x^2}{2\sigma_w^2}}}{\sqrt{2\pi}\sigma_w}\frac{32|x|^5}{\sigma_w^{10}}\\
\label{eq:h6}\left|\frac{\partial^6 K(x,a)}{\partial a^6}\right|&\leq h_6(x)=\frac{e^{-\frac{x^2}{2\sigma_w^2}}}{\sqrt{2\pi}\sigma_w}\frac{256|x|^6}{\sigma_w^{12}}
\end{align}
\fi
Clearly, the above functions are integrable since they are 
  found in the integrands of the central absolute moments of the Gaussian 
  distribution.
Therefore, conditions \ref{cond:dom_integrand} and \ref{cond:dom_deriv} of
  Lemma \ref{lemma:leibniz} are met by the integrand of \eqref{eq:kl_peak_power}
  and the integrand's derivatives.

The use of Lemma \ref{lemma:taylor} is conditional on the integrals over $x$
  of $K(x,a)$ and its derivatives in \eqref{eq:dKodd} and \eqref{eq:dKeven} 
  being continuous on $a\in[0,\sqrt{b}]$.
To prove the continuity of a function $f(x)$ on the interval $[u,v]$,
  it is sufficient to show that $\lim_{x\rightarrow x_0}f(x)=f(x_0)$ 
  for all $x_0\in[u,v]$.
We prove that $\int_{-\infty}^\infty K(x,a)dx$ is continuous as follows:
\begin{align}
\lim_{a\rightarrow a_0}\int_{-\infty}^\infty K(x,a)dx=\int_{-\infty}^\infty\lim_{a\rightarrow a_0}K(x,a)dx=\int_{-\infty}^\infty K(x,a_0)dx
\end{align}
where the first equality is due to the application of the dominated 
  convergence theorem, which is valid since we provide the 
  function $g(x)$ above that dominates $K(x,a)$ and is integrable, 
  and the second equality is due to the continuity of $K(x,a)$.
Similar steps can be used to prove the continuity of the integrals of the
  derivatives of $K(x,a)$, with the ultimate result being the satisfaction
  of the continuity condition of Lemma \ref{lemma:taylor}.
  
\subsection{Using an $\mathcal{O}(\sqrt{n}\log n)$-bit secret}
\label{app:key}
Here we demonstrate how Alice and Bob can construct a binary
  coding scheme that, on average, requires an
  $\mathcal{O}(\sqrt{n}\log n)$-bit secret.
This is done in two stages.
First, Alice and Bob randomly select the symbol periods that they will use
  for their transmission by flipping a biased coin $n$ times, with probability
  of heads $\tau$ to be assigned later. 
The $i^{\text{th}}$ symbol period is selected if the $i^{\text{th}}$ flip is
  heads.
Denote the number of selected symbol periods by $\eta$ and note that
  $\expected{\eta}=\tau n$.
Alice and Bob then use the best public binary codebook with codewords of length
  $\eta$ on these selected $\eta$ symbol periods.
They also generate and share a random binary vector $\mathbf{k}$ where
  $p_{\mathbf{K}}(\mathbf{k})=\prod_{i=1}^{\eta}p_K(k_i)$ with
  $p_K(0)=p_K(1)=\frac{1}{2}$.
Alice XORs $\mathbf{k}$ and the binary representation of the codeword 
  $\mathbf{c}(W_k)$.
The symbol location selection is independent of both the symbol and 
  the channel noise.
When Alice is transmitting a codeword, the distribution of 
  each of Willie's observations is
  $\mathbb{P}_s=(1-\tau)\mathcal{N}(0,\sigma_w^2)+\frac{\tau}{2}\left(\mathcal{N}(-a,\sigma_w^2)+\mathcal{N}(a,\sigma_w^2)\right)$
and, thus,
\begin{align}
\label{eq:kl_tau}\mathcal{D}(\mathbb{P}_w\|\mathbb{P}_s)&=&\int_{-\infty}^{\infty}\frac{e^{-\frac{x^2}{2\sigma_w^2}}}{\sqrt{2\pi}\sigma_w}\ln\frac{e^{-\frac{x^2}{2\sigma_w^2}}/\sqrt{2\pi}\sigma_w}{\frac{(1-\tau)e^{-\frac{x^2}{2\sigma_w^2}}}{\sqrt{2\pi}\sigma_w}+\frac{\tau}{2}\left(\frac{e^{-\frac{(x+a)^2}{2\sigma_w^2}}}{\sqrt{2\pi}\sigma_w}+\frac{e^{-\frac{(x-a)^2}{2\sigma_w^2}}}{\sqrt{2\pi}\sigma_w}\right)}dx
\end{align}
There is no closed-form expression for \eqref{eq:kl_tau}, but we can upper-bound
  it using Lemma \ref{lemma:taylor}.
The Taylor series expansion with respect to $a$ around $a=0$ can be done using
  Lemma \ref{lemma:leibniz}, with conditions for Lemmas \ref{lemma:taylor}
  and \ref{lemma:leibniz} proven similarly as in Theorem 
  \ref{th:achievability_peak_power}.
This yields the following bound:
\begin{eqnarray}
\label{eq:tv_taylor_bound_tau}\mathcal{V}_T(\mathbb{P}_w^n,\mathbb{P}_s^n)&\leq& \frac{\tau a^2}{2\sigma_w^2}\sqrt{\frac{n}{2}}
\end{eqnarray}
The only difference in (\ref{eq:tv_taylor_bound_tau}) from 
  (\ref{eq:tv_taylor_bound_pp}) is $\tau$ in the numerator.
Thus, if Alice sets the product $\tau a^2\leq\frac{cf(n)}{\sqrt{n}}$, with 
  $c$ and $f(n)$ as previously defined, she
  limits the performance of Willie's detector.
This product is the average symbol power used by Alice. 
Now fix $a$ and set $\tau=\mathcal{O}(1/\sqrt{n})$.
Since, on average, $\tau n$ symbol periods are selected, it takes 
  (again, on average) $\mathcal{O}(\sqrt{n})$ 
  positive integers to enumerate the selected symbols.
There are $n$ total symbols, and, thus, it takes at most $\log(n)$ bits to 
  represent each selected symbol location and $\mathcal{O}(\sqrt{n}\log n)$ 
  bits to represent all the locations of selected symbols.
Also, the average length of the secret binary vector $\mathbf{k}$ is 
  $\mathcal{O}(\sqrt{n})$ bits.
Thus, on average, Alice and Bob need to share $\mathcal{O}(\sqrt{n}\log n)$ 
  secret bits for Alice to reliably transmit $\mathcal{O}(\sqrt{n})$ bits in
  $n$ LPD channel uses employing this coding scheme.

\ifx\techreport\undefined
\else
\subsection{Proof of the generalized version of Fact \ref{fact:totvar_testperf}}
\label{app:bayes}
\begin{IEEEproof}[Proof (Fact \ref{fact:totvar_testperf_bayes})]
Upon observing $x$, Willie's hypothesis test selects either the 
  null hypothesis $H_0$ or the alternate hypothesis $H_1$.
Denote by $p_0(x)=p(x|H_0)$ and $p_1(x)=p(x|H_1)$ the probability density 
  functions 
  of $x$ conditioned on either hypothesis $H_0$ or $H_1$ being true;
  $p_0(x)$ and $p_1(x)$ are therefore the probability density functions of
  $\mathbb{P}_0$ and $\mathbb{P}_1$.
Denote by $p(H_0|x)$ and $p(H_1|x)$ the probabilities of hypotheses
  $H_0$ and $H_0$ being true conditioned on the observation $x$.
Since the optimal hypothesis test uses the maximum \emph{a posteriori} 
  probability rule, the
  probability $\mathbb{P}_c$ of Willie's optimal test being correct, averaged 
  over all observations, is as follows:
\begin{align}
\mathbb{P}_c&=\int_{\mathcal{X}}\max(p(H_0|x),p(H_1|x))p(x)dx\\
\label{eq:bayes_pc3}&=\int_{\mathcal{X}}\max(\pi_0p_0(x),\pi_1p_1(x))dx
\end{align}
where $\mathcal{X}$ is the support of $p_0(x)$ and $p_1(x)$, and
  \eqref{eq:bayes_pc3} follows from Bayes' theorem.
Let $\mathbb{P}_e=1-\mathbb{P}_c=1-\int_{\mathcal{X}}\max(\pi_0p_0(x),\pi_1p_1(x))dx$ denote the error probability of Willie's 
  optimal test.
Now, since $\max(a,b)=\frac{a+b+|a-b|}{2}$, $\mathbb{P}_e$ can be expressed as 
  follows:
\begin{align}
\label{eq:bayesints}\mathbb{P}_e&=1-\frac{1}{2}\left(\pi_0\int_{\mathcal{X}}p_0(x)dx+\pi_1\int_{\mathcal{X}}p_1(x)dx\right)-\frac{1}{2}\int_{\mathcal{X}}|\pi_0p_0(x)-\pi_1p_1(x)|dx\\
\label{eq:bayesPe2}&=\frac{1}{2}-\frac{1}{2}\|\pi_0p_0(x)-\pi_1p_1(x)\|_1
\end{align}
where \eqref{eq:bayesPe2} is due to the probability densities integrating to one
  over their supports in the first two integrals of \eqref{eq:bayesints},
  $\pi_0+\pi_1=1$, and the last integral in \eqref{eq:bayesints} being the
  $\mathcal{L}_1$ norm.
We can lower-bound \eqref{eq:bayesPe2} using the triangle inequality for the
  $\mathcal{L}_1$ norm:
\begin{align}
\mathbb{P}_e&\geq\frac{1}{2}-\frac{1}{2}\left(\|\pi_0p_0(x)-\pi_0p_1(x)\|_1+\|\pi_0p_1(x)-\pi_1p_1(x)\|_1\right)\\
\label{eq:bayesPe3}&=\frac{1}{2}-\frac{|\pi_0-\pi_1|}{2}-\frac{\pi_0}{2}\|p_0(x)-p_1(x)\|_1
\end{align}
where \eqref{eq:bayesPe3} follows from the $\mathcal{L}_1$ norm of a probability
  density function evaluating to one and $\pi_0>0$.
If $\pi_1>\pi_0$, the following application of the triangle inequality
  yields a tighter bound:
\begin{align}
\mathbb{P}_e&\geq\frac{1}{2}-\frac{1}{2}\left(\|\pi_1p_1(x)-\pi_1p_0(x)\|_1+\|\pi_1p_0(x)-\pi_0p_0(x)\|_1\right)\\
\label{eq:bayesPe4}&=\frac{1}{2}-\frac{|\pi_0-\pi_1|}{2}-\frac{\pi_1}{2}\|p_0(x)-p_1(x)\|_1
\end{align}
By Definition 1, 
  $\frac{1}{2}\|p_0(x)-p_1(x)\|_1=\mathcal{V}_T(\mathbb{P}_0,\mathbb{P}_1)$.
Since $\min(a,b)=\frac{a+b-|a-b|}{2}$, we can combine \eqref{eq:bayesPe3} and
  \eqref{eq:bayesPe4} to yield
\begin{align}
\label{eq:bayesPe}\mathbb{P}_e&\geq\min(\pi_0,\pi_1)-\max(\pi_0,\pi_1)\mathcal{V}_T(\mathbb{P}_0,\mathbb{P}_1)
\end{align}
which completes the proof.
\end{IEEEproof}
\fi

\bibliographystyle{IEEEtran}
\bibliography{IEEEabrv,../../../papers}
\end{document}